\documentclass[10pt,conference]{IEEEtran}
\usepackage{amsmath,amssymb,amsfonts,bm} 
\usepackage{algorithmic}
\usepackage{algorithm}
\usepackage{array}
\usepackage{caption}
\usepackage{subcaption}
\usepackage{textcomp}
\usepackage{stfloats}
\usepackage{bigints}
\usepackage{url}
\usepackage{verbatim}
\usepackage{graphicx}
\usepackage{mathtools}
\usepackage{dsfont}
\usepackage{esvect}
\usepackage{accents}
\usepackage{cite}
\usepackage{pbalance}
\usepackage{xcolor}
\newcolumntype{M}[1]{>{\centering\arraybackslash}m{#1}}
\newcommand*{\QEDA}{\null\nobreak\hfill\ensuremath{\blacksquare}}%

\newcommand\myeqa{\mathrel{\stackrel{\makebox[0pt]{\mbox{\normalfont\scriptsize (a)}}}{=}}}
\newcommand\myeqb{\mathrel{\stackrel{\makebox[0pt]{\mbox{\normalfont\scriptsize (b)}}}{=}}}

\begin{document}

\title{Joint Communication and Sensing in Aerial Corridors: A Novel Stochastic Geometry Framework}
% note positions of commas and nonbreaking spaces ( ~ ) LaTeX will not   brea k

\author{Harris K.~Armeniakos$^*$, 
        Petros S. Bithas$^{**}$,
        Athanasios G.~Kanatas$^*$,
        and Harpreet S. Dhillon\IEEEauthorrefmark{2} \\
       $^*$Department of Digital Systems, University of Piraeus, Piraeus, Greece \\
       $^{**}$Department of Digital Industry Technologies, National and Kapodistrian University of Athens, Greece \\
        \IEEEauthorrefmark{2}Bradley Department of Electrical and Computer Engineering, Virginia Tech, Blacksburg, VA, 24061, USA\\
        Email: \{harmen,kanatas\}@unipi.gr, pbithas@dind.uoa.gr, hdhillon@vt.edu
        % <-this % stops a space
%\thanks{The authors are with the Department
%of Digital Systems, University of Piraeus, 18534 Piraeus, Greece (e-mail:
%\{harmen,kanatas\}@unipi.gr).}% <-this % stops a space
}
\maketitle

\begin{abstract}
In unmanned aerial vehicle (UAVs) networks, joint communication and sensing (JCAS) is emerging as a key enabler to support extended and continuous communication and sensing capabilities for sixth-generation (6G) services among UAVs operating in swarms. Within this framework, aerial corridors provide a structured environment for supporting coordinated and reliable operations. In this paper, a comprehensive framework is presented to investigate the JCAS coverage probability (CP) of a UAV-base station (BS) in an aerial corridor populated by UAV-user equipments (UEs). The corridor is modeled as a finite cylinder, within which a fixed number of UAV-UEs are spatially distributed according to a three-dimensional (3D) binomial point process (BPP). The UAV-BS is assumed to be equipped with a realistic 3D third generation partnership project (3GPP) antenna pattern and exploits radar sensing capabilities to track a known UAV-UE. Subsequently, the tracked UAV-UE is assumed to perform uplink communication with the UAV-BS. Accordingly, the JCAS CP at the UAV-BS is analyzed under the presence of clutter and uplink communication interference, and exact-form analytical expressions are derived. To the best of our knowledge, this is the first work to develop a stochastic geometry framework for the rigorous analysis of JCAS performance in aerial corridors, with UAV-UE locations modeled as a 3D BPP. Among several insights, results show that increasing the directivity of the UAV-BS antenna beams leads to notable JCAS performance gains, particularly for shorter UAV corridors.
\end{abstract}


% \fi
\IEEEpeerreviewmaketitle

\section{Introduction}
The integration of unmanned aerial vehicles (UAVs) into sixth-generation (6G) wireless networks is expected to enable a wide range of communication and sensing applications. In various use cases, UAVs are considered to operate cooperatively in swarms to serve ground devices for delivery services \cite{alqudsi2025uav}, communication or energy transfer \cite{swarms}. In addition, UAV-to-UAV (U2U) communication networks are important because they allow direct coordination among drones without relying on ground infrastructure, which is essential for swarm operation, collision avoidance, relaying, and real-time data exchange. Consequently, U2U communication networks have attracted considerable research attention, e.g., \cite{Azari2020U2UCellular,KarimiBidhendi2024UAVCorridors,Prabhath2025Spectrum3D}. In particular, in \cite{Azari2020U2UCellular}, direct U2U links are analyzed coexisting with the cellular uplink, and underlay and overlay spectrum sharing is compared under realistic channel, antenna, and power-control assumptions. In \cite{KarimiBidhendi2024UAVCorridors}, it is shown that as UAV traffic becomes concentrated along aerial corridors, cellular networks should be optimized for reliable connectivity across the corridor through a suitable antenna-tilt and transmit-power design. In \cite{Prabhath2025Spectrum3D},
a framework is developed to evaluate spectral efficiency and spectrum utilization in 3D aerial networks, highlighting the importance of interference management and efficient spectrum use in multi-UAV systems. As the number of UAVs increases, the management of UAV traffic has gained  interest from various authorities, such as the Federal Aviation Administration and the National Aeronautics and Space Administration (NASA) \cite{FAA}. In particular, similar to ground vehicles and piloted aircraft, as UAVs become more common, their movement might be limited to specific aerial highways, also known as \emph{UAV corridors}, which should also comply with air traffic regulation authorities \cite{cherif20213d,karimi2024optimizing}. The main objectives of these air corridors are to minimize collision risks, avoid disruption to existing air traffic, and facilitate integration with the current airspace management system. With most UAVs traveling along corridors, network operators might now focus on providing reliable connectivity within
these specific areas rather than throughout the sky \cite{10974465}. Although the research trend for UAV network design is already moving towards the concept of aerial corridors \cite{10974465, 11363236, karimi2024optimizing, 10711847}, the development of air corridors is still in an early stage, as various aspects should be decided, such as optimization of design-related issues \cite{10974465} and performance criteria.

As another key enabling technology in next-generation wireless networks, joint communication and sensing (JCAS) offers a sustainable way to jointly support communication and sensing functionalities. With 3GPP having included the support of UAV-BSs as part of the fifth generation (5G), JCAS technology can now be utilized from UAV-BSs to monitor and manage swarms of UAV-UEs. Subsequently, the collected JCAS data can be forwarded to terrestrial BSs \cite{10098686, 10004900}. In this context, a particularly appealing application is UAV tracking, where a UAV-BS monitors the position and trajectory of a UAV-UE while also supporting its uplink communication. However, while the UAV-UEs swarm may communicate with other UAV-BSs, severe interference can arise and cause both clutter and communication interferences \cite{10098686, 10004900}. To address this challenge, appropriate antenna configuration and beam design at the UAV-BS are crucial for effective interference mitigation. Nevertheless, while JCAS has been extensively investigated in UAV-assisted networks \cite{10879807, 10977743}, its application to UAV networks remains largely unexplored  \cite{10004900}.

Stochastic geometry provides a powerful mathematical framework for modeling the spatial aspects of UAV networks and analyzing their performance. When a fixed and predetermined number of UAVs are deployed in a finite region, their spatial distribution can be well described by a simple but reasonable homogeneous BPP. In the context of UAVs-assisted communications, a one-dimensional (1D) BPP was proposed in \cite{10711847}, for the purpose of analytical tractability, and experimentally verified in \cite{11363236} to model the spatial locations of UAVs in corridors modeled as finite line segments. Nevertheless, the 3D BPP has already been exploited \cite{9250029} for modeling the locations of UAVs in a vertically-located cylinder, where the length of the cylinder was assumed to determine the deployment height of the UAVs. However, in UAV networks, the investigation of JCAS performance necessitates the statistical characterization of not only the Euclidean distances between the UAV-BS and the UAV-UEs, but also the statistical characterization of the angular distances. The latter arise naturally if the UAV-BS is equipped with a 3D antenna pattern.  \emph{To the best of the authors’ knowledge, this is the first work that rigorously investigates the manifestation of JCAS technology in aerial networks, where UAV-UEs are modeled as a 3D BPP in aerial cylinder corridors.}

\emph{Contributions}: A comprehensive framework is presented to investigate the JCAS performance of a UAV-BS in a UAV network. By employing the 3D BPP to model the spatial locations of UAV-UEs, a fixed number of UAV-UEs equipped with isotropic antennas are deployed in an aerial corridor modeled as a finite cylinder. The UAV-BS is assumed to be equipped with a realistic 3D 3GPP-based antenna pattern and exploits beamforming capabilities to monitor and track a UAV-UE while the tracked UAV-UE is assumed to communicate with the UAV-BS. Subsequently, exact-form analytical expressions for JCAS coverage probability (CP) are derived. Since the inclusion of the 3GPP antenna requires the statistical characterization of angular coordinates, the joint probability density function (PDF) of the 3D spatial location of a random UAV-UE in the corridor is derived as a key intermediate result in the JCAS CP analysis.

\section{System Model}
\subsection{Spatial and Network Modeling}
Consider a JCAS-enabled UAV network that integrates UAV-UEs and a UAV-BS. Assume that the UAV-UEs hover in a UAV aerial corridor, as will be discussed in the following, with rules enforced by appropriate traffic authorities. The UAV corridor is modeled as a finite \emph{3D cylinder}, which has been proposed as the most reasonable geometric model to approximate a real UAV corridor \cite{yu2025hybrid}. Consequently, we assume that the UAV-UEs are independently and uniformly distributed within the finite cylindrical region $\mathcal{C}(L,R)$, where $\mathcal{C}(L,R)$ represents a 3D cylinder centered at the origin $\mathbf{o}$, with axial extent from $-L/2$ to $L/2$ and radius $R$. Therefore, a fixed number of $N$ UAV-UEs equipped with isotropic antennas  assumed to be distributed within $\mathcal{C}(L,R)$ and their spatial locations form a 3D BPP $\Phi$. Although isotropic antennas are assumed for the UAV-UEs, consistent with their typically limited directional capabilities \cite{10467182}, the proposed framework can be readily extended to incorporate directional UAV-UE antennas. A UAV-BS equipped with a 3D 3GPP-based antenna pattern is assumed to be centered at the origin $\mathbf{o}$, as shown in Fig. 1. In this context, let $\{\mathbf{w}_i\} \equiv \Phi$ denote the 3D spatial locations of the UAV-UEs given by $\{\mathbf{w}_i\}=\{(x_i,y_i,z_i)\}_{i=1:N}$. Then, let $r_i = \| \mathbf{w}_i \|$, for $1 \leq i \leq N$, denote the Euclidean distance between the $i$-th UAV-UE and the UAV-BS. For the adopted scheduling mechanism, it is assumed that sensing and communication functionalities are performed using different time-frequency resources, which is a common assumption. The exploitation of different scheduling mechanisms is left for future work. The UAV-BS enables radar sensing capabilities to track a UAV-UE's movement trajectory, while the tracked UAV-UE performs an uplink communication with the UAV-BS. 

\begin{figure}[!t]
    \centering
    \includegraphics[keepaspectratio,width=3.47in]{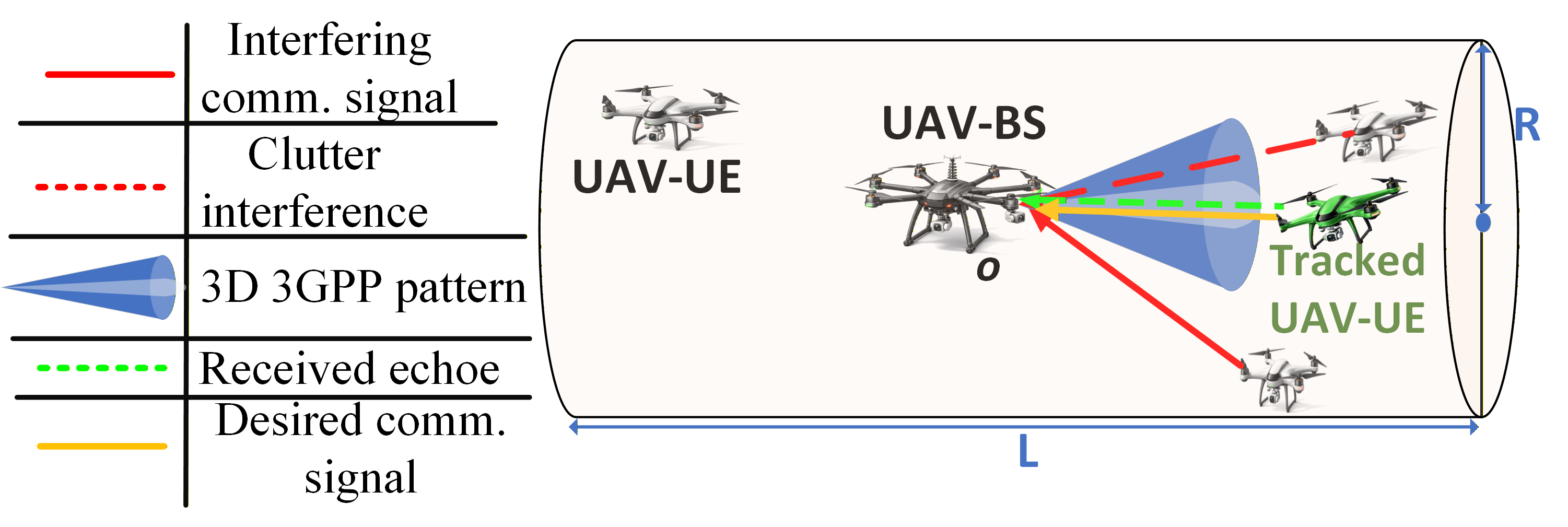}
       \caption{Illustration of the UAV JCAS-enabled network with a UAV-BS equipped with a 3D 3GPP-based antenna pattern and $N$ UAV-UEs forming a 3D BPP in a UAV corridor.}
    \label{fig:Fig2}
\end{figure}

%In this work, a worst-case scenario interference scenario is considered in the mmWave network and therefore, the maximum gain of the BSs’ antennas is assumed to be always directed towards the receiver

%The UAV-BS is assumed to exploit beamforming techniques to enable JCAS with the target UAV-UE.

\subsection{Beamforming and Antenna Modeling}
 The UAV-BS's antenna array is able to produce a directive beam both for the sensing and the communication functionalities.  A realistic 3GPP-based antenna pattern \cite{3GPP_TR_38_900_14_2_0} is adopted, and thus the actual antenna gain within the 3dB beamwidth range is not constant. When directed towards $(\theta_0,\phi_0)$, this antenna pattern has  vertical,  horizontal, and  3D radiation power patterns (in dB) given by
\begin{align}\label{eq6}
       &G_{3gpp}^V(\theta,\theta_0) = -{\rm{min}}\Bigg\{ 12\Bigg(\frac{\theta-\theta_0}{\theta_{3dB}} \Bigg)^2, SLA \Bigg\}, \\ \nonumber
       &G_{3gpp}^H(\phi,\phi_0) = -{\rm{min}}\Bigg\{ 12\Bigg(\frac{\phi-\phi_0}{\phi_{3dB}} \Bigg)^2, SLA \Bigg\},  \\ \nonumber
       &G_{3gpp}^{3D}(\theta,\theta_0,\phi,\phi_0) =\\ \nonumber
       &G_{{\rm{max}}}-{\rm{min}}\Big\{-G_{3gpp}^V(\theta,\theta_0) -G_{3gpp}^H(\phi,\phi_0)\Big\} \nonumber,     
 \end{align}
 where $\theta_{3dB}$ and $\phi_{3dB}$ are the vertical and horizontal $3$~dB beamwidths and $SLA = 30$~dB is the front-to-back ratio. The multiplicative
antenna gain factor is denoted by $g_{3gpp}^{3D}(\theta,\theta_0,\phi,\phi_0)$ and the multiplicative maximum antenna gain factor is denoted by $g_{{\rm{max}}}$. The UAV-BS takes advantage of the sensing data gathered from the tracked UAV-UE to infer its instantaneous position/trajectory and updates the beam steering direction accordingly \cite{10806828}. More specifically, the UAV-BS is assumed to steer its beam toward the tracked UAV-UE. This reflects the practical JCAS operation, in which beam alignment is performed using angular estimation. Consequently, the tracked UAV-UE continuously experiences the maximum antenna gain, while the other UAV-UEs are seen through off-boresight gains acting as clutter interference.

\subsection{Path Loss and Channel Models}
A key characteristic of UAV networks is the ability of UAVs to establish line-of-sight (LoS) links above rooftop level. In this context, following the standard power-law path-loss model for the path between a UAV-UE and the UAV-BS, the random path loss function is denoted by $l(\|\mathbf{w}_i\|) = r_i^{-\alpha}$, with $\alpha$ denoting the path loss exponent for the LOS propagation links. The amplitudes of the communication and sensing channels are assumed to experience Nakagami-$m$ fading. Therefore, the power gains of the channels $h_{i}^j$ between the $i$-th UAV-UE and the UAV-BS follow the gamma distribution, while $j \in \{c,s\}$ is used to characterize the (independent) channels during the communication and detection phase. Therefore, $h_i^j \sim {\rm{Gamma}}\big(m,\frac{1}{m}\big)$ with the shape and scale parameters of $h_i^j$ being $m$ and $1/m$, respectively.

\subsection{Association Policy and SIR Definition}
\subsubsection{UAV-UE Association Policy}
In this work, the tracked UAV-UE is chosen uniformly at random among the set of $N$ available UAV-UEs and tracking functionality is captured over time through its different spatial realizations. Subsequently, the tracked UAV-UE performs uplink communication with the UAV-BS. Once the location $\mathbf{w}_0$ of the tracked UAV-UE is determined, let $\Phi^! = \Phi \setminus \{\mathbf{w}_0\}$.

\subsubsection{Communication and Sensing SIR Definition}
The UAV-BS measures the reflected echo from the tracked UAV-UE whereas at the same time  suffers from clutter interference caused by the rest $(N-1)$ UAV-UEs in the corridor. By following the radar equation, the  sensing SIR from the tracked UAV-UE received at the UAV-BS's can be written as 
\begin{equation}
{\rm{SIR}}_s = \frac{P_{\rm{echo}}}{I_s} = \frac{g_{{\rm{max}}}^2 h_{0}^s \sigma_0 l(||\mathbf{w}_0||)^2}{\sum\limits_{i \in \Phi^!} g_{3gpp}^{3D}(\theta_i,\theta_{0},\phi_i,\phi_{0})^2 h_{i}^s \sigma_i  l(||\mathbf{w}_i||)^2}, 
\end{equation}
where $P_{\rm{echo}} = p_t g_{{\rm{max}}}^2 h_{0}^s \sigma_0 l(||\mathbf{w}_0||)^2\frac{\lambda^2}{(4 \pi)^3}$ and $I_s$ denotes the aggregate clutter interference observed at the UAV-BS's given by  $I_s = \sum\limits_{i \in \Phi^!} p_t g_{3gpp}^{3D}(\theta_i,\theta_{0},\phi_i,\phi_{0})^2 h_{i}^s \sigma_i  l(||\mathbf{w}_i||)^2\frac{\lambda^2}{(4 \pi)^3}$. In particular, $g_{3gpp}^{3D}(\theta_i,\theta_{0},\phi_i,\phi_{0})$ denotes the gain of the UAV-BS antenna in the direction $(\theta_{i},\phi_{i})$ of the $i$-th UAV-UE causing clutter interference, when the UAV-BS antenna is directed toward the tracked UAV-UE at $(\theta_{0},\phi_{0})$. Moreover, $p_t$ denotes the equal transmit power for all UAV, $\lambda$ denotes the carrier wavelength, and $\sigma_0$ denotes the random radar cross-section (RCS) area of the tracked UAV-UE. By adopting the Swerling I model, the PDF of RCS  $\sigma_i$ for a random UAV-UE is given by $f_{\sigma_i}(\sigma) = \frac{1}{\overline{\sigma}}\exp(-\sigma/\overline{\sigma})$, where $\overline{\sigma}$ denotes the average RCS. 

The tracked UAV-UE is assumed to perform uplink communication with the UAV-BS while the UAV-BS suffers from interference caused by the rest of the $(N-1)$ active UAV-UEs\footnote{An assumption adopted in this work is that the interfering UAV-UEs in the communication phase belong to a different BPP, denoted by $\Psi$, than those involved in the sensing/tracking phase. This assumption is made solely for analytical tractability. In the more general and realistic case where the interferer locations across the sensing and communication phases are dependent, the corresponding aggregate interference terms become coupled, and the analysis would require conditioning on the entire BPP $\Phi$. This would result in a substantially different and much more involved mathematical treatment, which lies beyond the scope of this work.}. The latter captures a worst-case interference scenario. The communication SIR from the $i$-th tracked UAV-UE received at the UAV-BS's antenna array is given by
\begin{equation}
{\rm{SIR}}_c = \frac{S_0}{I_c}=\frac{g_{{\rm{max}}} h_{0}^c l(||\mathbf{w}_0||) }{\sum\limits_{i \in \Psi} g_{3gpp}^{3D}(\theta_i,\theta_{0},\phi_i,\phi_{0}) h_{i}^c l(||\mathbf{w}_i||)},
\end{equation}
where  $I_c$ denotes aggregate uplink interference from the
UAV-UEs to the UAV-BS \cite{binti2022qoe}.

\section{JCAS Performance Analysis}

\emph{Definition 1:} \emph{The JCAS CP $\mathcal{P}_{cs}(\theta_c, \theta_s)$, is defined as the probability that the received SIR of the communication and sensing functions at the receiving UAV-BS are simultaneously not less than a predefined threshold $\theta_c$ and $\theta_s$ required for successful communication and tracking, respectively, i.e., $\mathcal{P}_{cs}(\theta_c, \theta_s)  \triangleq \mathbb{P}[ {\rm{SIR}}_{c}>\theta_c, {\rm{SIR}}_s>\theta_s]$. }

\subsection{Sensing Coverage Probability}
The statistical characterization of both the sensing and communication CP depends on both the Euclidean distances and the angular coordinates of the UAV-UEs. Therefore, the joint PDF of the 3D spatial location of a UAV-UE in terms of the spherical coordinates is first derived. 

\emph{Lemma 1:} \emph{Let $\mathbf{p}_i$ denote the 3D spatial location of the $i$-th UAV-UE in terms of the spherical coordinates $(r,\theta,\phi)$ in $\mathcal{C}(L,R)$. The joint PDF $f_{\mathbf{p}_i}(r,\theta,\phi)$ is given by }
\begin{equation}
f_{\mathbf{p}_i}(r,\theta,\phi) = \frac{r^2 \sin(\theta)}{V},
\end{equation}
\emph{where $V=\pi R^2L$ and $\theta \in[0,\pi]$, $\phi \in [-\pi,\pi]$,  $r \in [0,r_{\max}(\theta,\phi)]$ and $r_{\max}(\theta,\phi) = \min\Big\{  \frac{L/2}{|\sin(\theta)\cos(\phi)|},\frac{R}{\sqrt{1-\sin^2(\theta)\cos^2(\phi)}} \Big\}$.}

\textit{Proof.} Refer to Appendix A. \QEDA

Conditioned on the location $\mathbf{p}_0$ of the tracked UAV-UE, the conditional Laplace transform of the aggregate clutter interference power distribution can now be derived.

\begin{figure*}[!b]
\hrulefill
\begin{equation} \tag{5}  \label{eqLTs}
\mathcal{L}_{I_s}(s|\mathbf{p}_0) = \Big[\int_{0}^{\infty}\iiint_{(r_i,\theta_i,\phi_i)}   \Big(1+\frac{s \sigma_i g_{3gpp}^{3D}(\theta_i,\theta_{0},\phi_i,\phi_{0})^2    l(||\mathbf{w}_i||)^2}{m}\Big)^{-m} f_{\mathbf{p}_i}(r_i,\theta_i,\phi_i) f_{\sigma_i}(\sigma_i) {\rm{d}}  r_i {\rm{d}} \theta_i {\rm{d}} \phi_i  {\rm{d}}  \sigma_i \Big]^{N-1}
\end{equation}
\begin{equation} \tag{7} \label{eqLTc}
 \mathcal{L}_{I_c}(s|\mathbf{p}_0) = \Big[\iiint_{(r_i,\theta_i,\phi_i)}   \Big(1+\frac{s g_{3gpp}^{3D}(\theta_i,\theta_{0},\phi_i,\phi_{0})    l(||\mathbf{w}_i||)}{m}\Big)^{-m} f_{\mathbf{p}_i}(r_i,\theta_i,\phi_i)  {\rm{d}}  r_i {\rm{d}} \theta_i {\rm{d}} \phi_i  \Big]^{N-1}
\end{equation}
\end{figure*}

\emph{Lemma 2:} \emph{Conditioned on the location  $\mathbf{p}_0$ of the tracked UAV-UE, the conditional Laplace transform  $\mathcal{L}_{I_s}(s|\mathbf{p}_0)$ of the aggregate clutter interference power distribution is given by \eqref{eqLTs}, shown at the bottom of the page, where $r_i= ||\mathbf{w}_i||$ and $f_{\mathbf{p}_i}(r_i,\theta_i,\phi_i)$ is given by Lemma 1.}

\textit{Proof.} Refer to Appendix B. 

\emph{Proposition 1:} \emph{The conditional sensing CP $\mathcal{P}_{s}(\theta_s|\mathbf{p}_0)$ of the UAV-BS is given by}
\begin{equation} \tag{6}
\begin{split}
\mathcal{P}_{s}(\theta_s|\mathbf{p}_0) = &\int_{0}^{\infty} \sum_{k=0}^{m - 1}  \frac{{(-m \theta_s)}^k }{ (\sigma_0 g_{{\rm{max}}}^2 l^2(||\mathbf{w}_0||))^k k!} \\
&\times {\Bigg[ \frac{\partial^{k} \mathcal{L}_{I_s}(s|\mathbf{p}_0)}{\partial{s^k}}\Bigg]}_{s_0}f_{\sigma_0}(\sigma_0) {\rm{d}}  \sigma_0,  
\end{split}
\end{equation}
\emph{for $s_0 = \frac{m \theta_s}{\sigma_0 g_{{\rm{max}}}^2l^2(||\mathbf{w}_0||)}$ and $||\mathbf{w}_0|| = r_0$}.

\textit{Proof.} The proof for deriving Proposition 1 follows conceptual lines similar to those presented in \cite[Theorem 1]{TWC} and thus is omitted here for brevity.  \QEDA

\subsection{Communication Coverage Probability}

\emph{Lemma 3:} \emph{The conditional Laplace transform  $\mathcal{L}_{I_c}(s|\
\mathbf{p}_0)$ of the aggregate interference power distribution is given by \eqref{eqLTc}, shown at the bottom of the page.}

\textit{Proof.} The proof for deriving Lemma 3 follows a similar procedure as the proof for deriving Lemma 2. \QEDA

% Now, the conditional communication coverage probability conditioned on  $\mathbf{p}_0$ is derived.

\emph{Proposition 2:} \emph{The conditional communication CP $\mathcal{P}_{s}(\theta_s|\mathbf{p}_0)$ of the UAV-BS is given by}
\begin{align} \tag{8}
\mathcal{P}_{c}(\theta_c|\mathbf{p}_0) = &\sum_{k=0}^{m - 1}  \frac{{(-m \theta_c)}^k }{( g_{{\rm{max}}} l(||\mathbf{w}_0||))^k k!} {\Bigg[ \frac{\partial^{k} \mathcal{L}_{I_c}(s|\mathbf{p}_0)}{\partial{s^k}}\Bigg]}_{s_0}, \nonumber 
\end{align}
\emph{for $s_0 = \frac{m \theta_c}{ g_{{\rm{max}}} l(||\mathbf{w}_0||)}$ and $||\mathbf{w}_0|| = r_0$}.

\textit{Proof.} The proof for deriving Proposition 2 follows conceptual lines similar to those of the proof presented in \cite[Theorem 1]{TWC}. \QEDA

\subsection{JCAS Coverage Probability}
\emph{Theorem 1:} \emph{The JCAS CP of a UAV-BS in a UAV corridor-assisted JCAS-enabled  network is given by}
\begin{align} \tag{9}
 \mathcal{P}_{cs}(\theta_c, \theta_s) &=   \iiint_{(r_0,\theta_0,  \phi_0)}  \mathcal{P}_{c} (\theta_c|\mathbf{p}_0)  \mathcal{P}_{s}(\theta_s|\mathbf{p}_0) \\\nonumber
 &\times f_{\mathbf{p}_0}(r_0, \theta_0,  \phi_0)   {\rm{d}} r_0 {\rm{d}} \theta_0 {\rm{d}} \phi_0. \nonumber
\end{align}

\textit{Proof.} Recalling Definition 1, the JCAS CP is given by
\begin{equation} \tag{10} \label{proofTh1}
\begin{split}
&\mathcal{P}_{cs}(\theta_c, \theta_s) = \mathbb{E}[\mathds{1}({\rm{SIR}}_{c} \geq \theta_c) \times \mathds{1}({\rm{SIR}}_{s} \geq \theta_s)] \\
&\myeqa  \mathbb{E}_{\mathbf{p}_0}[\mathds{1}({\rm{SIR}}_{c} \geq \theta_c | \mathbf{p}_0) \times \mathds{1}({\rm{SIR}}_{s} \geq \theta_s |\mathbf{p}_0] \\
&  \myeqb \mathbb{E}_{\mathbf{p}_0}[\mathbb{P}({\rm{SIR}}_{c} \geq \theta_c |\mathbf{p}_0) \times \mathbb{P}({\rm{SIR}}_{s} \geq \theta_s |\mathbf{p}_0)],
\end{split}
\end{equation}
where (a) follows directly from the independence of the point process of interfering UAV-UEs $\Phi^!$ and $\Psi$ for sensing and communication, respectively, (b) follows after applying the indicator function which results to be the product of the conditional coverage probabilities that have been proved in Proposition 1 and Proposition 2. After substituting Proposition 1 and Proposition 2 in \eqref{proofTh1} and applying the expectation on \eqref{proofTh1} to jointly decondition over the location of the tracked UAV-UE $\mathbf{p}_0$ w.r.t $f_{\mathbf{p}_0}(r_0, \theta_0,  \phi_0)$, Theorem 1 is obtained. \QEDA

\emph{Remark:} \emph{The results obtained in Proposition 1 and Proposition 2 can be exploited to directly investigate the marginal sensing and coverage probabilities, respectively, for a fixed or random 3D position of the tracked UAV-UE. This is a first step towards the investigation of simpler yet more tractable frameworks that enable the derivation of closed-form expressions for JCAS coverage probability.}

\section{Numerical Results \& Discussions}
In this section, representative numerical results are presented to evaluate the performance achieved in the proposed  UAV corridor-assisted JCAS-enabled network due to space constraints. The accuracy of the analytical results is verified by comparing them with the simulated ones obtained using Monte Carlo simulations. The following parameters have been used, unless stated otherwise: $L = 100$~m, $R=5$, $\alpha = 2.2$, $m = 2$,  $N = 10$,  $(\theta_c,\theta_s) = (-10,-10)$~dB, $(\phi_{3dB},\theta_{3dB}) = (45^o,45^o)$, $G_{\rm{max}}=8$~dBi,  and $\overline{\sigma} = 0.7 m^2$.

\begin{figure}[!t]
    \centering
    \includegraphics[keepaspectratio,width=3.45in]{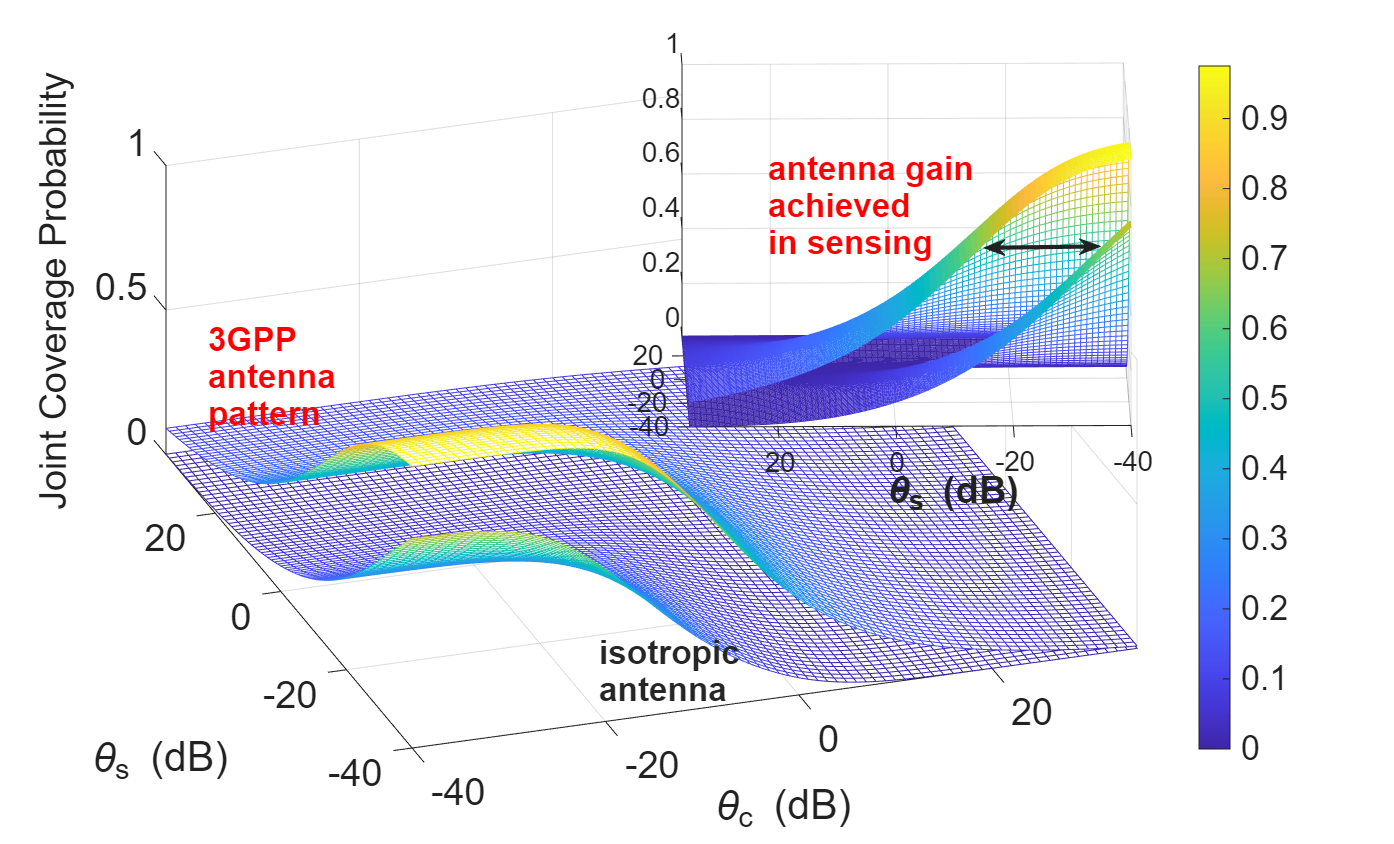}
       \caption{JCAS CP versus $\theta_c$ and $\theta_s$.}
    \label{fig:Fig2}
\end{figure}

Fig. \ref{fig:Fig2} presents the  analytical JCAS CP, obtained using Theorem 1, versus the SIR thresholds $\theta_c$ and $\theta_s$ to capture and demonstrate the impact of sensing and communication functionalities on the JCAS performance. The JCAS CP under an isotropic antenna with $G_{3D}^{\rm{iso}} = G_{\rm{max}}$ is also simulated for comparison purposes and to highlight the impact of a directional pattern. The first observation is that the sensing CP degrades more smoothly than the communication CP. This can be attributed to the distinct form of the sensing SIR, which captures the round-trip propagation of the radar signal via the corresponding path-loss and antenna-gain terms, along with the contribution of the RCS. Moreover, it is observed that the gain offered by the 3GPP UAV-BS's antenna array in the representative case of the sensing performance is notable, even if a wide 3dB beamwidth is assumed.    

\begin{figure}[!t]
    \centering
    \includegraphics[keepaspectratio,width=3.0in]{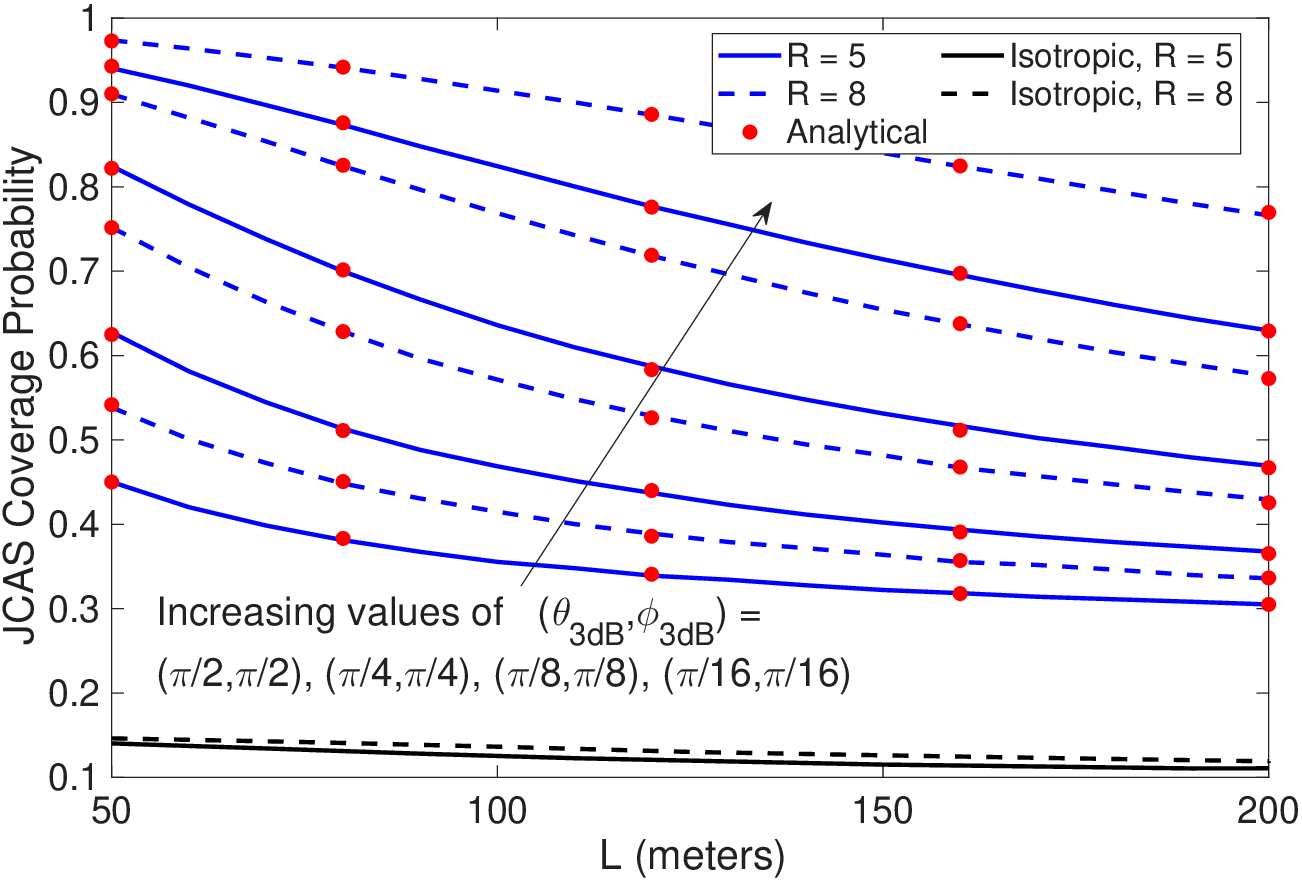}
       \caption{JCAS CP versus $L$ for different values of 3dB beamwidths and for two values of UAV corridor radius $R$.}
    \label{fig:Fig3}
\end{figure}

\begin{figure}[!t]
    \centering
    \includegraphics[keepaspectratio,width=3.0in]{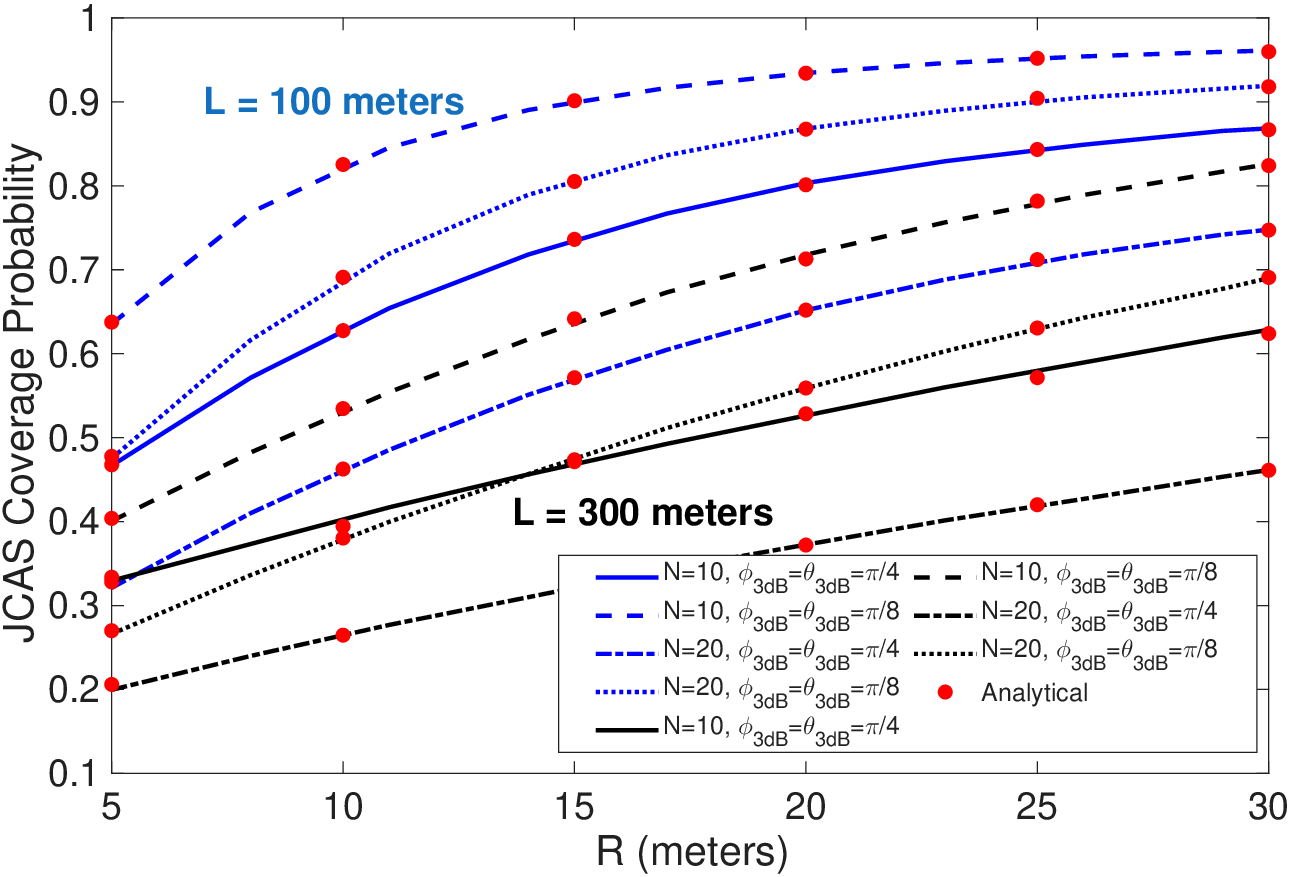}
       \caption{JCAS CP versus $R$ for different values of $\phi_{3dB}, \theta_{3dB}$ and $N$ given two values of UAV corridor length $L$.}
    \label{fig:Fig4}
\end{figure}

Fig. \ref{fig:Fig3} presents the  analytical (denoted by markers) and simulated JCAS CP versus $L$ for different values of 3dB beamwidths and $R$. The JCAS CP under an isotropic antenna with $G_{3D}^{\rm{iso}} = G_{\rm{max}}$ is also simulated for completeness. A key observation is that the JCAS CP is maximized for UAV corridors with large radius and small lengths. This is beacuase of the fact that as the deployment radius of the UAV corridor increases, the interfering UAV-UEs become more likely to lie outside the mainlobe of the UAV-BS antenna pattern, thus leading to a reduction in the aggregate interference power. Note that as the beams become more directional, a notable improvement in JCAS performance is achieved for the shorter UAV corridor lengths, independently of the values of $R$. For larger corridor lengths, a significant improvement in performance is observed only when more directional beams are employed and/or larger values of $R$ are considered. Clearly, even the wider beams achieve a notable increase in JCAS performance as compared to the case with isotropic antennas, which highlights the necessity of designing directional beams to suppress interference in JCAS-enabled UAV networks.

Fig. \ref{fig:Fig4} presents the  analytical (denoted by markers) and simulated JCAS CP versus $R$ for different values of 3dB beamwidths and $L$ as well as the number of UAV-UEs. Interestingly, for longer UAV corridors, e.g., $L=300$~m, the performance improvement due to more directional UAV-BS beams is more significant when the corridor radius $R$ is larger. This observation indicates that in large-scale UAV corridors, increased beam directivity is particularly effective in suppressing interference. By contrast, for shorter UAV corridors, this trend is weaker and is not consistently observed for both values of $N$. Moreover, for shorter UAV corridors, the increase in JCAS CP with an increase of $R$ is not uniform. Indeed, the improvement is more pronounced when $R$ varies from small to moderate values, whereas it becomes weaker for larger values of $R$. This is because enlarging a narrow corridor substantially reduces dominant interfering UAV-UEs, which are close to the UAV-BS and inside the mainlobe of the UAV-BS's antenna. Once these dominant interference conditions have been greatly alleviated, a further increase in $R$ provides only a small additional benefit, until eventually it saturates.

%Finally, it can be easily noticed that the JCAS coverage performance of a FA system clearly outperforms the corresponding one under the single-port antenna system even for the smaller values of FA size.

% By assuming that a UAV-BS equipped with a 3D 3GPP antenna exploits JCAS capabilities, the performance of the  JCAS CP is analytically investigated.

\section{Conclusion}
This paper developed a novel stochastic geometry framework for analyzing the JCAS performance of UAV swarms operating within aerial corridors. Among several key system-level insights, it was shown that: i) as the beams of the UAV-BS's antenna pattern become more directional, a notable increase in JCAS performance is achieved for the smaller lengths of UAV corridors, and ii) for longer UAV corridors, the performance enhancement achieved as the beams of the UAV-BS's antenna pattern  become more directional is more pronounced for the larger values of radius $R$. The results highlight the challenges associated with interference mitigation and motivate the need to optimize key design parameters of UAV corridors to optimize JCAS performance in UAV networks. The demonstrated benefits of directional beamforming in mitigating interference from surrounding UAVs may also be relevant to counter-UAS surveillance, where a UAV of interest must be reliably tracked in dense aerial traffic. Extending the proposed framework to unknown and non-cooperative UAVs is therefore an interesting direction for future work.

\appendices
\section{Proof of Lemma 1}
The joint PDF $f_{\mathbf{w}_i}(x,y,z)$ of the Cartesian coordinates of a random UAV-UE in a cylinder $\mathcal{C}$ of volume $V$ is given by $f_{\mathbf{w}_i}(x,y,z) = \frac{1}{V}$, where $\mathcal{C} = \Big\{(x,y,z): x \in [-\frac{L}{2},\frac{L}{2}] ,\quad y^2+z^2\leq R^2\Big\}$. Applying the transformation of the changing coordinate system to $(r,\theta,\phi)$, yields $\iiint_{(r,\theta,\phi)} \frac{1}{V} r^2 \sin\theta {\rm{d}}r {\rm{d}}\phi {\rm{d}}\theta = 1$, where $f_{\mathbf{p}_i}(r,\theta,\phi) = \frac{r^2 \sin\theta}{V}$. Next, from the definition of $ \mathcal{C}$, 
\begin{equation}  \tag{11} \label{eq_proof2}
    |x|\leq L/2 \rightarrow r|\sin\theta\cos\phi| \leq L/2 \rightarrow r \leq \frac{L/2}{|\sin\theta\cos\phi|},
\end{equation} 
\begin{align}  \tag{12} \label{eq_proof3}
    &y^2 + z^2  = r^2 (\sin^2\theta\sin^2\phi + \cos^2\theta) \\
    &\myeqa r^2 (1-\sin^2\theta\cos^2\phi) \leq R^2 \rightarrow r\leq\frac{R}{\sqrt{1-\sin^2\theta\cos^2\phi}}, \nonumber
\end{align}
where (a) follows from $ \underbrace{\sin^2\theta\cos^2\phi}_{x} +\underbrace{\sin^2\theta\sin^2\phi}_{y} +  \underbrace{\cos^2\theta}_{z} = 1$. Now, the maximum radius $r_{\rm{max}}(\theta,\phi)$ yields the minimum value among \eqref{eq_proof2} and \eqref{eq_proof3} and therefore $r_{\max}(\theta,\phi) = \min\Big\{  \frac{L/2}{|\sin(\theta)\cos(\phi)|},\frac{R}{\sqrt{1-\sin^2(\theta)\cos^2(\phi)}} \Big\}$.

\section{Proof of Lemma 2}
The Laplace transform of $I_s$ conditioned on $\mathbf{p}_0$  is given by 
\eqref{eqLTproof}, shown at the bottom of the page, where (a) follows from the  moment generating function (MGF) of $h_{i}^s$  (b) follows from the identically and independently distributed (i.i.d.)  $N-1$ random variables of the interfering reflections and after averaging over $\sigma_i$.  Lemma 2 is obtained after deconditioning on $f_{\mathbf{p}_i}(r_i,\theta_i,\phi_i)$ for  $(N-1)$ interfering UAV-UEs. 

\begin{figure*}[!b]
\hrulefill
\begin{equation} \label{eqLTproof}  \tag{13}
\begin{split}
&\mathcal{L} _{I_s}(s|\mathbf{p}_0)=\mathbb{E}_{I_s}[e^{-sI_s}]=\mathbb{E}_{\Phi^{!},h_{i}^s,\sigma_i}\Big[e^{-s \sum_{i \in \Phi^{!}}  \sigma_i h_{i}^s g_{3gpp}^{3D}(\theta_i,\theta_{0},\phi_i,\phi_{0})^2    l(||\mathbf{w}_i||)^2}\Big|\mathbf{p}_0\Big] \\
&\myeqa \mathbb{E}_{\Phi^{!},\sigma_i}\Bigg[\prod_{i \in \Phi^{!}}\Bigg[  \Big(1+\frac{s\,  \sigma_i  g_{3gpp}^{3D}(\theta_i,\theta_{0},\phi_i,\phi_{0})^2    l(||\mathbf{w}_i||)^2  )}{m} \Big)^{-m} \Big|\mathbf{p}_0 \Bigg] \Bigg] \\
&\myeqb \Bigg[\mathbb{E}_{r_i=||\mathbf{w}_i||,\theta_i,\phi_i}\Bigg[ \int_{0}^{\infty}\Big(1+\frac{s\,  \sigma_i  g_{3gpp}^{3D}(\theta_i,\theta_{0},\phi_i,\phi_{0})^2    l(||\mathbf{w}_i||)^2  )}{m} \Big)^{-m} f_{\sigma_i}(\sigma_i) {\rm{d}}  \sigma_i \Big|\mathbf{p}_0 \Bigg] \Bigg]^{N-1}, \\
\end{split}
\end{equation}
\end{figure*}

\section*{Acknowledgment}
This work was co-funded by the University of Piraeus Research Center (UPRC) and the EU HORIZON programme under iSEE-6G GA No. 101139291. H. S. Dhillon gratefully acknowledges the support of US National Science Foundation (Grants CNS-2107276 and CNS-2225511).

\bibliographystyle{IEEEtran}
\bibliography{IEEEabrv,references}
\balance

\end{document}